\documentclass[preprint,12pt,sort&compress]{elsarticle}

\usepackage{graphicx}
\usepackage{amssymb}
\usepackage{amsmath}

\makeatletter
\def\ps@pprintTitle{%
   \let\@oddhead\@empty
   \let\@evenhead\@empty
   \let\@oddfoot\@empty
   \let\@evenfoot\@oddfoot
}

\journal{Journal Name}

\begin{document}

\begin{frontmatter}


\title{A Dynamic Niche Model for the Emergence and Evolution of Mutualistic Network Structures}





\author[a]{Weiran Cai\corref{cor1}} 
\ead{wrcai@ucdavis.edu}
\author[a,b]{Jordan Snyder} 
\author[c,d]{Alan Hastings}
\author[a,d]{Raissa M. D'Souza} 
\address[a]{Network Theory Group, Department of Computer Science, University of California at Davis, 1 Shield Ave., Davis, CA 95616}
\address[b]{Graduate Group in Applied Mathematics, UC Davis, CA 95616}
\address[c]{Department of Environmental Science and Policy, UC Davis, CA 95616}
\address[d]{Santa Fe Institute, 1399 Hyde Park Rd, Santa Fe, NM 87501}

\cortext[cor1]{Corresponding author}



\begin{abstract}
Mutualistic interactions are vital constituents of ecological and socio-economic systems. Empirical studies have found that the patterns of reciprocal relations among the participants often shows the salient features of being simultaneously nested and modular. Whether and how these two structural properties of mutualistic networks can emerge out of a common mechanism however remains unclear. We propose a unified dynamic model based on the adaptation of niche relations that gives rise to both structural features. We apply Hutchinson's concept of niche interaction to networked cooperative species. Their niche relation evolves under the assumption of fitness maximization. Modularity and nestedness emerge concurrently through the accumulated local advantages in the structural and demographic distribution. A rich ensemble of key dynamical behaviors are unveiled in the dynamical framework. We demonstrate that mutualism can exhibit either a stabilizing or destabilizing effect on the evolved network, which undergoes a drastic transition with the overall competition level. Most strikingly, the adaptive network may exhibit a profound nature of history-dependency in response to environmental changes, allowing it to be found in alternative stable structures. The adaptive nature of niche interactions, as captured in our framework, can underlie a broad class of ecological relations and also socio-economic networks that engage in bipartite cooperation.
\end{abstract}


\end{frontmatter}


\medskip
Mutualism, as evidenced in plant-pollinator, seed-disperser or mycorrhizal systems, typically exhibits a pattern of ordered collective interactions, which may be depicted by a nontrivial network representation. Such networks are comprised of bipartite guilds of multiple species (plants and animals for example), which are typically involved in both cross-guild cooperation and within-guild competition. The most distinctive properties shared by these networks are the modular and nested patterns in the cooperative interspecific relation that are consistently over-expressed relative to their randomized counterparts \cite{Bascompte2003,Olesen2007,Toju2014,Bascompte2013,Bronstein2015}. While a modular organization implies that most mutual links can be enclosed in several clusters, a nested structure indicates that more often than not the partners of one species of a lower degree (specialist) are a subset of the partners of another species of a higher degree (generalist). Despite the statistical prevalence of these two features, little consensus has been reached on the role that mutualism has on the dynamical properties, ranging from network stability to the impacts of environmental changes \cite{Okuyama2008,Bastolla2009,Thebault2010,Saavedra2011,James2012,Allesina2012,Grilli2016}. The uncertainty is largely due to the lack of understanding of the dynamical origin of network formation. Here, we demonstrate a unified principle that explains how both global patterns of mutualistic networks can emerge from localized interactions. 

Various classes of models have been proposed to explore the origin of mutualistic network structures \cite{Guimaraes2007a,Valverde2018,Sole-Ribalta2018,Suweis2013,Zhang2011,Saavedra2009}. Among these, two seemingly contrary principles are prevalent. Models based on adaptive population dynamics envision that all species are initially equivalent and develop a non-trivial interaction structure solely due to stochastic fluctuations under an overall incentive \cite{Suweis2013,Zhang2011}. For instance, the pursuit of maximal individual abundances may guide the evolution of interspecific relations, which provides a profound explanation for the emergent nestedness; yet modularity is left unexplained. In contrast, models based on the concept of specific niches assume that species are endowed with distinguishable traits and that network features arise exclusively out of static niche relations \cite{Saavedra2009,Williams2000,Cattin2004}. Indeed, the crucial role of niche relationships in the formation of network structure has been demonstrated in various empirical analyses \cite{Olesen2007,Santamaria2007,Guimaraes2007,Ollerton2006}. However, the static niche framework captures only the snapshot of an evolved network, disregarding the underlying driving mechanism. The specificity of niche relations is indeed the consequence of adaptation and continues to evolve \cite{Armbruster1998,Ollerton2006,Vazquez2009,Olesen2009}. Thus, a unified dynamic niche model, incorporating both population dynamics and niche relations, is desired. 

Under the assumption of maximizing individual fitnesses, mutualistic species evolve their mutually linking pattern to an optimal structure based on niche interactions. To formulate this as a dynamical process, we apply Hutchinson's concept of niche adaptation to a network of cooperative species incorporating niche dynamics \cite{Hutchinson1957,Hutchinson1978,MacArthur1967,Cohen1978}. In the following sections, we begin by demonstrating how modularity and nestedness can emerge simultaneously through a positive feedback of local advantages in the structural and population distributions. We then delve into the dynamical properties of the evolved network by focusing on its resilience. We show that mutualism can have either a stabilizing or destabilizing effect on the evolved network, undergoing a drastic transition between the two behaviors with the intensity of within-guild competition. At a large evolutionary time scale, we illustrate that the interspecific linking pattern may exhibit a prominent hysteresis in response to environmental changes. The intrinsic history-dependency means the mutualistic network may remain in alternative structures even if the original external condition is recovered.

\begin{figure}[h!]
\hskip -0.6cm
  \includegraphics[width=14.5cm]{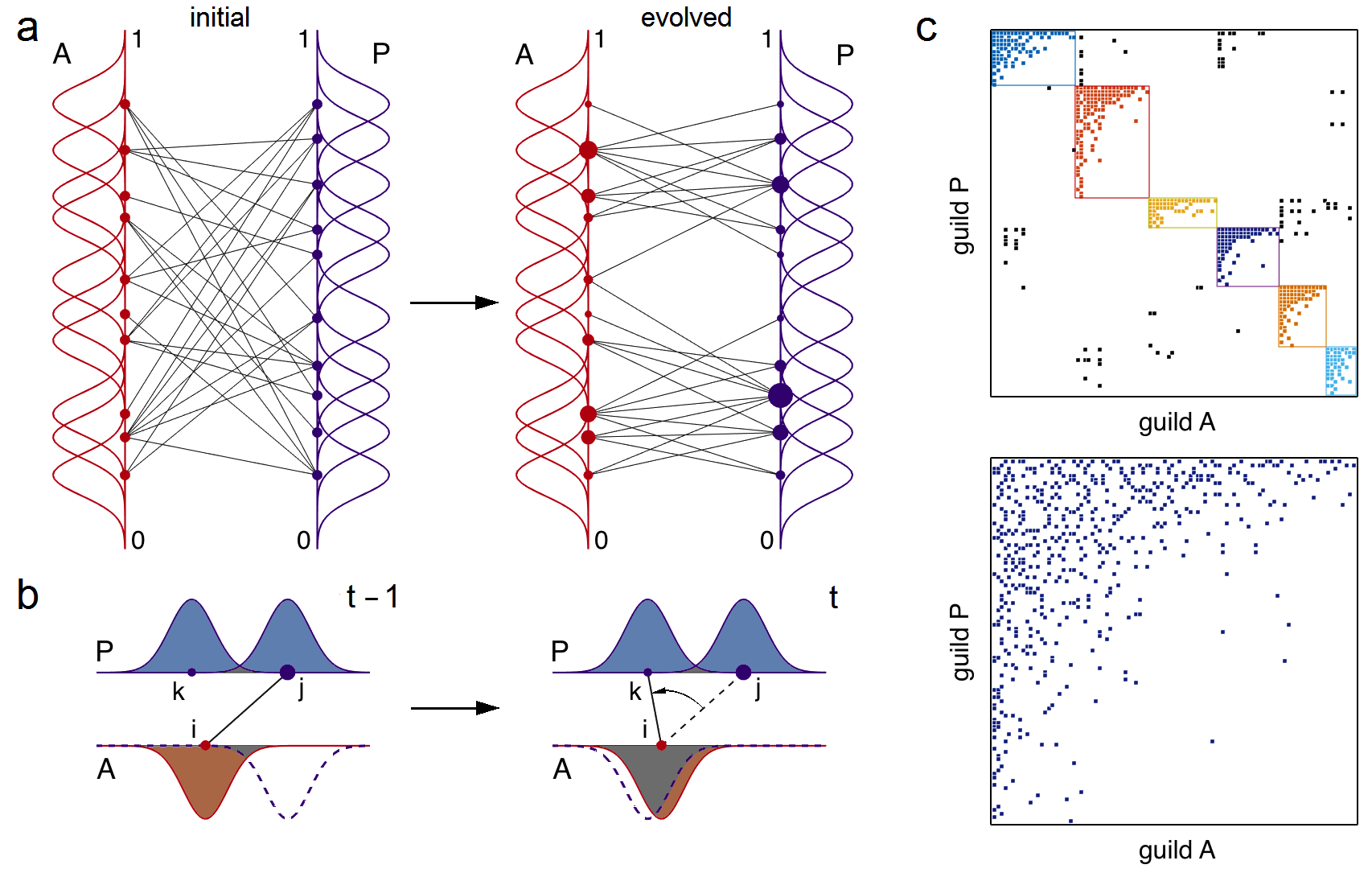}\\
  \caption{Dynamic niche model. \textbf{a}, Adaptation of niche relations. Two mutually interacting guilds (A and P) of species evolve from an arbitrary topology (left) to a structured one of stable partnership (right). Each species $i$ is endowed with a niche profile: a Gaussian function $H_i(s)$ with the center $\bar{s_i}$ chosen randomly on the niche axis $[0,1]$. Its abundance, represented by a disk of proportional size, is governed by population dynamics (Eq.~\ref{eq:lv}). An example network of 20 species is illustrated. \textbf{b}, Edges rewire over time to maximize their individual fitnesses. The example demonstrates the rewiring of a node $i$ from a high-abundant partner $j$ to a new one $k$ with a higher niche overlap.
The gray areas show the niche overlaps that determine the within- and cross-guild interaction strengths $\beta_{kj}$ and $\gamma_{ij}$. \textbf{c}, Typical network structure observed in the 
evolved steady state. The adjacency matrix is reordered to emphasize the modular (upper panel) and nested (lower panel) structures of the same network. Note the nested structure is also embedded within each module as seen in the upper panel.
The example evolved network, with 100 species in each guild, shows significantly higher modularity and nestedness ($Q=0.6620$, $N=0.9574$) than the randomized networks ($P\le 0.0001$), simulated here for $\Omega_m=0.05$ and $\Omega_c=0.1$.}\label{fig_def}
\end{figure}

\section{Dynamic Niche Model}
We consider a network comprised of multiple species in two distinct guilds ($A$ and $P$ in analogy with animals and plants), which are involved simultaneously in mutualistic interactions with selected partner species in the opposite guild and subject to competition with all rival species within their own guild. The connectivity and coupling strengths are encoded in coupling matrices $\{\gamma_{ij}\}$ and $\{\beta_{ij}\}$ for mutualistic and competitive interactions, respectively. We assume that each species possesses two fundamental characteristics: its niche and its abundance (see Materials and Methods). The niche profile for species $i$ is given by a Gaussian function $H_i(s)$, representing its statistical distribution
on a one-dimensional niche axis $[0,1]$ (see Fig. \ref{fig_def}a). We assume that the center positions $\bar{s_i}$ of the niche functions are randomly sampled from the niche axis and do not change over time (for discussion of evolution of niche positions, see SI). We further assume that if two species interact, their coupling strength is proportional to their niche overlap \cite{MacArthur1967,Scheffer2006}. The species abundances $\{n_i\}$ follow the generalized Lotka-Volterra population dynamics, where the mutualistic interactions are described by Holling type II functional response \cite{Bastolla2009}. At fixed time intervals, a randomly chosen species attempts to rewire to a different mutualistic partner in the opposite guild (Fig. \ref{fig_def}b) in order to maximize its own abundance \cite{Suweis2013}. The time intervals are chosen to be sufficiently long to guarantee that the population dynamics reaches an equilibrium between the rewiring attempts.

\begin{figure}[t!]
\hspace{0.1cm}
  \includegraphics[width=13cm]{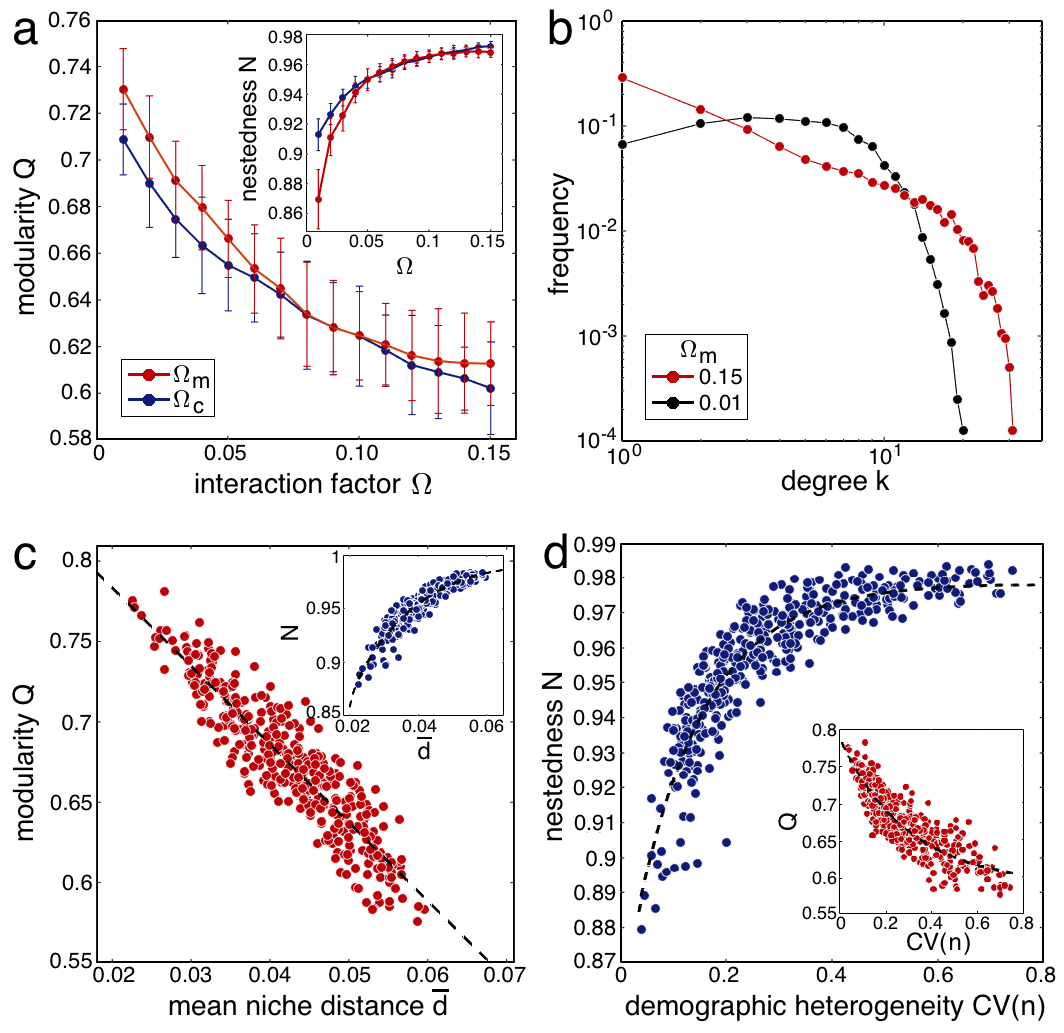}
  \caption{Evolved network structure. \textbf{a}, Structural measures versus interaction factors. Modularity $Q$ (nestedness $N$ in inset) decreases (increases) with either mutualistic or competitive factor ($\Omega_m$ or $\Omega_c$). The error bars represent one standard deviation. \textbf{b}, Emergent degree distribution. The distribution shows a typical truncated power law for a high mutualistic factor (shown for $\Omega_m = 0.15$) while it approaches a narrower single-peaked distribution when $\Omega_m$ decreases ($\Omega_m = 0.01$). \textbf{c}, Modularity and nestedness (in inset) versus average niche distance $\bar{d}$ of linked species. Modularity decreases linearly while nestedness increases with $\bar{d}$ (fitted by an exponential curve). \textbf{d}, Coherent relation between network structure and demographic distribution. Nestedness $N$ and modularity $Q$ (inset) show a monotonic positive and negative correlation with the demographic heterogeneity $CV(n)$. The data points for both relations are fitted by exponential curves. The scatter plot consists of 400 realizations, which are generated for randomly chosen interaction factors $\Omega_m$ and $\Omega_c$. }\label{fig_IF}
\end{figure}

\section{Evolved Structure and Niche Relation}
With these primitive assumptions in place, our numerical simulations generate a rich ensemble of network structures, exhibiting a broad range of modularity and nestedness. The nestedness is measured by the nestedness temperature coefficient (NTC) \cite{Rodriguez-Girones2006} and the modularity is calculated with the leading eigenvector algorithm \cite{Newman2006}. An example evolved network demonstrates highly modular and nested structures as shown by the reordered adjacency matrices in the upper and lower panels of Fig. \ref{fig_def}c, respectively. In all simulations we start from a randomly connected mutualistic network $\{\gamma_{ij}\}_{t=0}$ with a specified connectance (link density) $C_0$, and the network evolves as described above until reaching a steady state where all macroscopic structural and demographic measures remain approximately constant with time, which we refer to as the evolved steady state. 

Most prominently, a number of modules emerge with nested link patterns embedded in all of them (exemplified in the upper panel of Fig. \ref{fig_def}c). Such structure is typically found in empirical seed-disperser or plant-pollinator networks. Species in these modules are more densely connected to the local hubs (generalists) than to species in other modules. Furthermore, except for a few specialists, most of them belong persistently to only one of the modules once the evolved steady state is reached. In this sense, the entire community has settled into a macroscopic order after exploring a landscape of numerous possibilities of niche relations \cite{Levin1998}. 

The global interconnection pattern is a manifestation of local interactions, which are regulated by the external environment. We mimic the overall environmental influence by changing the factors $\Omega_m$ and $\Omega_c$ for mutualistic and competitive interactions, respectively, which are the interaction intensity per unit of niche overlap (see Materials and Methods for definition). Contrary to existing models of mutualism, where the network structure is independent of the dynamics thereon, such change in the local interaction factors substantially alters the evolved structure (Fig. \ref{fig_IF}a). Concretely, we found that enhancing either the overall mutualistic or competitive intensity, by $\Omega_m$ or $\Omega_c$, contributes positively to the nestedness $N$ while suppresses the modularity $Q$. The two contrasting interactions thus regulate the network structure in a highly similar manner. Notably, the within-guild competition, which has so far been underestimated \cite{Gracia-Lazaro2018}, acts as a crucial determinant for the cross-guild partnership.

The degree distribution, describing the heterogeneity in the numbers of partners per species, evolves from a Poisson distribution of the initial random network to a typical truncated power law when both interaction intensities (in terms of $\Omega_m$ and $\Omega_c$) are high (Fig. \ref{fig_IF}b). When either intensity is reduced, the distribution however turns into a narrow single-peaked distribution simply due to the topological constraint:
Species contained within a smaller-sized module tend to possess comparable numbers of partners, which forces the degree distribution to be more homogeneous.

The evolved modular and nested structure reflects the niche affinity among partner species, as has been long suggested by empirical studies \cite{Olesen2007,Santamaria2007,Guimaraes2007}. We generate an ensemble of networks by randomly varying the interaction factors $\Omega_m$ and $\Omega_c$, mimicking a range of external conditions (see SI). We define the pair-wise niche distance as the separation of niche centers $d_{ij}=|\bar{s_i}-\bar{s_j}|$. The average niche distance $\bar{d}$, over all connected species pairs, is found to be negatively correlated with the modularity $Q$ (linear fitting) and positively correlated with the nestedness $N$ (exponential fitting), as shown in Fig. \ref{fig_IF}c. Hence, a more nested network structure can tolerate interspecific partnerships with less niche affinity (higher $\bar{d}$), while a more modular structure (smaller sizes of modules) is packed with species of more complementary niches (lower $\bar{d}$).

\section{Feedback of Local Advantages} 

The network structure and the demographic distribution of species abundances co-evolve to be correlated at the steady state. This is again demonstrated by generated networks for randomly chosen interaction factors. Concretely, a positive and a negative correlation for the nestedness $N$ and modularity $Q$ are identified against the demographic heterogeneity $CV(n)$ of abundances, respectively (see Figure \ref{fig_IF}d). Here, we measure the relative unevenness across all species abundances by the coefficient of variation $CV(n)=\sigma(n)/\bar{n}$, with $\bar{n}$ and $\sigma(n)$ denoting the average abundance and the standard deviation over all species, respectively. The relative demographic heterogeneity is thus accurately conveyed in the evolved network structure. In contrast, no monotonic relation is identified between overall abundance of the community and the structural measures (see SI).

Heuristically, the modular and nested structure is formed through a positive feedback of local advantages in the structural and demographic distributions. Under the incentive of increasing individual fitnesses, a preponderance in the abundance of a certain species attracts more remote partner species on the niche axis, which in turn enhance its own abundance. Links thus aggregate around a small number of separate local hubs (generalists). Modularity and nestedness hence emerge concurrently. This process contrasts with the existing models that handle them with separate mechanisms \cite{Guimaraes2007a,Valverde2018,Sole-Ribalta2018,Suweis2013,Zhang2011,Grilli2016}. The dynamics belongs to a broad class of localized preferential attachment processes, whereby `the rich get richer' under the constraints on the potential linkage \cite{Albert2002,Montoya2006}. Such cumulative local advantage, prevailing in socio-economic systems, thus also underlies mutualistic interaction patterns.

\section{Transition in Network Stability} 
A mutualistic network reaches an asymptotically constant link structure and a {balanced} population distribution at the evolved steady state. It is necessary to examine whether this population distribution on the evolved network can withstand transitory external interference. Local stability considers the response to small perturbations on species abundances. Concretely, we characterize the stability of the evolved network by calculating the real part of the leading eigenvalue of the Jacobian of the population dynamics (Eq. \ref{eq:lv}), $S=-Re(\lambda)_{max}$ (see SI). Conventional models study the network stability based on a fixed network topology under all external conditions, even when the intensity of involved interactions has changed \cite{Okuyama2008,Thebault2010,Allesina2012,Staniczenko2013,Grilli2016}. However, an ecological network structure typically adapts to the environment and changes its stability in a systematic manner. Such structural adaptation is the foundation of our model. 

\begin{figure}[h!]
\hskip -0.72cm
  \includegraphics[width=1.08\textwidth]{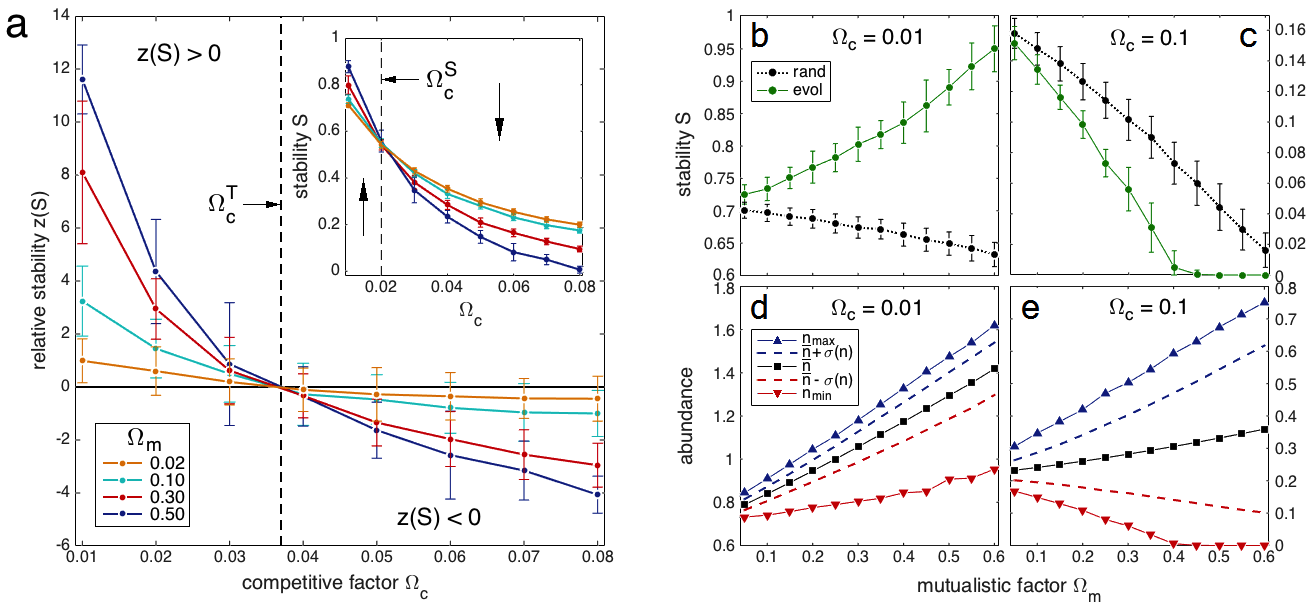}
  \caption{Role of competition on network stability. \textbf{a}, Transition of stability with competition factor. An evolved network is persistently more or less stable than the null model when $\Omega_c$ is below or above a transition point $\Omega_c^T (\approx 0.37)$, irrespective of the mutualistic factor $\Omega_m$. The relative stability is defined by the z-score $z(S)=(S-S_0)/\sigma_0$ of the stability of the evolved network, in comparison of that of the null model. Each $z(S)$ value is averaged over 50 evolved networks. The null model consists of 300 randomizations of the evolved network with preserved connectance and interaction factors. Inset: The tendency of network stability. Enhancing mutualistic interaction by $\Omega_m$ destabilizes the network in the presence of intensive competition ($\Omega_c>\Omega_c^S\approx 0.02$) while stabilizes it in the opposite case ($\Omega_c<\Omega_c^S$), as indicated by the arrows.
\textbf{b, c}, Opposite scenarios of network stability for two typical values of competition factor $\Omega_c$, separated by the transition point $\Omega_c^T$ or $\Omega_c^S$. The evolved network is more stable than the null model for $\Omega_c=0.01$ and is further stabilized by enhancing mutualistic factor $\Omega_m$ (panel b); the contrary case is shown for $\Omega_c=0.1$ (panel c). \textbf{d, e}, Changes in demographic distribution. The lower bound $n_{min}$ of the population shows opposite tendencies with the mutualistic factor $\Omega_m$ for the competition factor below or above $\Omega_c^S$.} \label{fig_stab}
\end{figure}

Most notably, we identify a drastic transition in the stability with the overall intensity of within-guild competition. This transition can be depicted by the relative stability $z(S)=(S-S_0)/\sigma_0$, which is the z-score of the local stability $S$ of the evolved network, with $S_0$ and $\sigma_0$ being the mean stability and standard deviation of an appropriate null model. The null model we use consists of an ensemble of randomized networks with the same connectance and interaction factors of the evolved network. An evolved network is persistently more or less stable than its random counterpart when the competition factor $\Omega_c$ is below or above a transition point $\Omega_c^T$, as shown in Fig. \ref{fig_stab}a. Note that the sign of the relative stability is irrespective of the mutualistic intensity $\Omega_m$, since all $z(S)$ curves intersect at the same $\Omega_c^T$. More explicitly, opposite scenarios are shown for the stability measures $S$, compared with that of the null model, for two typical values of $\Omega_c$ separated by the transition point $\Omega_c^T$ in Fig. \ref{fig_stab}b and \ref{fig_stab}c. Hence, whether a mutualistic network has a higher stability than the random network is predominantly determined by the overall intensity of the involved competition.

Moreover, the role of mutualism on stability is modulated by the competition intensity. Enhancing mutualistic interaction $\Omega_m$ may either stabilize or destabilize the network, depending on whether the competition factor $\Omega_c$ is below or above a threshold $\Omega_c^S$ ($< \Omega_c^T$; see inset of Fig. \ref{fig_stab}a). The opposite tendencies are again demonstrated for typical values of $\Omega_c$ in Fig. \ref{fig_stab}b and \ref{fig_stab}c. 

The dichotomy is consistent with the demographic distribution. Although the overall population of the community always increases with the mutualistic factor $\Omega_m$, the tendency of its lower bound $\min_i(n_i)$ depends crucially on the competition factor $\Omega_c$ (Fig. \ref{fig_stab}d and \ref{fig_stab}e). The network stability degrades or improves with the lower bound of the population, through their close relation $S=-Re(\lambda_m) \approx \min_i(n_i)$ (see \cite{Suweis2013} and SI). Beyond the threshold of competition intensity $\Omega_c^S$, enhancing mutualistic interaction can detriment the low-abundant species and thus degrade the stability. Hence, competition has a decisive impact on the network stability, which can even inverse the role of mutualistic interaction on it.

\section{Hysteresis in Structural Adaptation}
\begin{figure}[h!]
\hspace{-1.05cm}
  \includegraphics[width=1.08\textwidth]{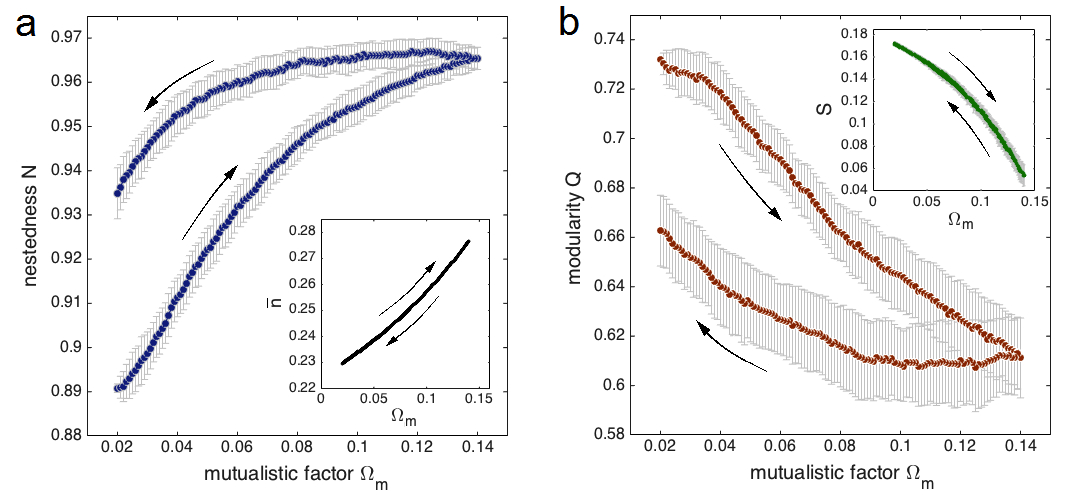}
  \caption{Hysteresis in structural adaptation. Nestedness (\textbf{a}) and modularity (\textbf{b}) show a strong history-dependence with the mutualistic factor $\Omega_m$. A very slow change rate ($\Delta\Omega_m=10^{-6}$ per time interval) is used when raising and restoring $\Omega_m$ so that the system is guaranteed to be always at quasi-steady-states. The network develops in preference in the direction of decreasing modularity and increasing nestedness, while it shows a strong resistance in the opposite direction. In parallel, the mean abundance $\bar{n}$ (inset of upper panel) and the measure of stability $S=-Re(\lambda)_{max}$ (inset of lower panel) are reversed along the same path when $\Omega_m$ is restored. The same population and stability level are recovered on the altered network structure. } \label{fig_irv}
\end{figure}

The network resilience is further examined under slow environmental changes, beyond the ecological time scale studied above. Such long-term changes may cause systematic alteration of the network that are irreversible even if the external environmental condition is restored \cite{Kauffman1993,Levin1998,Ollerton2006}. Numerically, we track the structural measures of an evolved network by raising and restoring the mutualistic factor $\Omega_m$, mimicking the impact of a gradually changing environment at an evolutionary time scale. A slow change rate of $\Omega_m(t)$ is used so that the entire process is guaranteed to be at quasi-steady-states. Strikingly, the trajectories of both modularity and nestedness values show an unclosed hysteresis with the control factor (see Fig. \ref{fig_irv}). 

This is essentially because the adaptation process proceeds preferably in the direction of merging (when $\Omega_m$ increases) rather than splitting modules (when $\Omega_m$ decreases). Species attach to the highly abundant generalists (on the ascending path) more easily than detach from them (on the descending path). The network thus shows a strong resistance to the decrease of modular sizes (increase of modularity $Q$). Consequently, species engaged in cooperation may adopt alternative niche relations even under the same environmental condition. Such hysteresis implies that the mutualistic network would freeze accidents of environmental history in its structure \cite{Kauffman1993,Levin1998}. 

However, this path-dependency affects only the structural properties. In the same process, the average community population $\bar{n}$ and the stability measure $S$ are well recovered along the same path when the influenced variable $\Omega_m$ is restored (see insets of Fig. \ref{fig_irv}). Hence, neither the overall population level nor local stability shows traces of the environmental change, despite that the underlying structure has been drastically altered. The same population level remains, along with the same stability, on a series of alternative network structures. Similar hysteresis phenomena are observable with other control parameters (see SI).

\section{Discussion} 

We have established that the nested and modular structural properties of mutualistic networks, which have previously been considered in separate mechanisms, may emerge from a common biological motivation. We have employed Hutchinson's principle that species interact via ecological niches in the pursuit of their individual fitnesses. The development of global interspecific linking pattern is a natural consequence of the adaptation of niche relations. Modularity and nestedness are cognate facets of an evolved complementary niche structure \cite{Fortuna2010}, which consolidates the local advantages via a positive feedback. These structural features are endogenously correlated with the niche affinity and demographic heterogeneity. Once such network structures arise, they not only constrain the mutualistic partnership, but profoundly influence the network's dynamical profile \cite{Levin1998}.

The adaptive nature of our model has revealed a rich ensemble of dynamical properties of mutualism that are unaccessible from the standpoint of assuming fixed network topologies, illuminating in particular the impact of mutualism on network resilience \cite{Okuyama2008,Thebault2010,Allesina2012,Staniczenko2013,Grilli2016}. Our analysis demonstrates the critical role played by competition on the network stability, which has hitherto been largely ignored \cite{Gracia-Lazaro2018}: whether a mutualistic network is more stable than its random counterpart is predominantly determined by the intensity of within-guild competition. Caution should be taken for a highly competitive community, where enhancing mutualism may even reduce the network resilience and harm species of lower abundances, which is reminiscent of the Matthew effect of cumulative advantage \cite{Merton1968}. On the other hand, the hysteresis of the adaptive structure strongly suggests that the interspecific relation in mutualistic networks shows a profound history-dependency in response to slow environmental changes and may possess alternative stable structures \cite{Kauffman1993,Levin1998}. It thus confirms the previous statistical studies signifying that historical factors are indispensable in the network formation \cite{Vazquez2009,Olesen2009}.

The proposed adaptation process has shown the applicability of Hutchinson's quintessential principle to networked cooperative species, which conveys a bottom-up perspective on mutualism. This principle, by simply incorporating niche overlaps, clarifies the common origin of heterogeneities exhibited in both structure and population \cite{Hutchinson1978,MacArthur1967,Scheffer2006}. Our framework has extended this concept from the continuous niche space, originally customized for niche competition, to the niche network. Such adaptive niche networks may underpin a broader class of fundamental relations and selection phenomena, such as host-parasitic and prey-predator interactions \cite{Warren2010,Williams2000,Cattin2004}. We anticipate that studies on random attacks and robustness \cite{Memmott2004}, invasion-extinction process \cite{Morris2003,Melbourne2008} and control of such adaptive networks \cite{Kiers2010} would be particularly revealing. This mechanism, with variations, can potentially be validated also for socio-economic cooperation networks \cite{May2008,Saavedra2009}.

\section*{Materials and Methods}{
We consider a bipartite network that contains interacting species in two guilds (denoted $A$ and $P$, in analogy with animals and plants). Each species $i$ is assigned a niche profile, which is formulated as a Gaussian function $H_i(s)$ with uniform width $\sigma$ and its center position $\bar{s_i}$ randomly chosen from the interval $[0,1]$ on a niche axis (see SI). Each species is involved in cross-guild mutualistic interactions with selected partner species (represented by a matrix $\{\gamma_{ik}\}$ between $A$ and $P$), in addition to competitive interactions with all rival species in its own guild ($\{\beta_{ij}\}$ for $A$ or $P$). We define the niche overlap $H_{ij}$ of a pair of species $i$ and $j$ as
\begin{equation}
\label{eq:h}
H_{ij}=\int H_i(s)H_j(s)ds=\exp\left(-\frac{\left(\bar{s_i}-\bar{s_j}\right)^2}{4\sigma^2}\right).
\end{equation}
which is the joint occupation probability of two species on the niche axis \cite{MacArthur1967,Scheffer2006}. The intensity of either type of pair-wise interaction is assumed to be proportional to the niche overlap $H_{ij}$. Specifically, 
\begin{subequations}
\label{eq:niche_overlap}
\begin{align}
     \text{mutualistic:}\quad \gamma_{ik}&=\Omega_m\cdot \theta_{ik}\cdot H_{ik}\\
     \text{competitive:}\quad \beta_{ij}&=
     \begin{cases}
     1, &i = j\\
     \Omega_c\cdot H_{ij}, &i\not=j     
     \end{cases}     
\end{align}
\end{subequations}
where $i,j\in G=(A$ or $P$) and $k\in \bar G=(P$ or $A$). $\{\theta_{ik}\}$ is the adjacency matrix, with the entries equal to $1$ if $i$ and $k$ interact, and $0$ if not. The proportionality coefficients $\Omega_m$ and $\Omega_c$ are the interaction factors for mutualistic and competitive interactions, respectively, which capture the overall environmental influence.

The species abundances evolve according to a set of Lotka-Volterra equations with Holling-Type II mutualistic functional response \cite{Bascompte2013}:
\begin{subequations}
\label{eq:lv}
\begin{align}
\frac{d n_i^A}{d t} &= n_i^A\left(\rho_i^A  - \sum\limits_j\beta_{ij}^A n_j^A + \frac{\sum\limits_k\gamma_{ik}^{AP}n_k^P}{1 + h\sum\limits_k \theta_{ik}^{AP}n_k^P}\right) \\ 
\frac{d n_i^P}{d t} &= n_i^P\left(\rho_i^P  - \sum\limits_j\beta_{ij}^P n_j^P + \frac{\sum\limits_k\gamma_{ik}^{PA}n_k^A}{1 + h\sum\limits_k \theta_{ik}^{PA}n_k^A}\right)
\end{align}
\end{subequations}
where the coupling strengths $\{\gamma_{ik}\}$ and $\{\beta_{ij}\}$ are defined above and updated during the evolution as described next. 

All species are assigned uniform abundances $n_i=n_0$ and connected to partner species across the guilds uniformly at random with a specified connectance $C_0$ in the initial condition. The system evolves according to Eq. \ref{eq:lv} and at each time interval $t=mT$ ($m$ is a positive integer), a species $i$ is chosen uniformly at random and one of its existing links $\gamma_{ij}$ is rewired to a randomly selected different mutualistic partner species $j'$ with probability $p_{ij}$. $T$ is chosen to be sufficiently large to guarantee that the population dynamics reaches equilibrium between subsequent rewiring attempts. At the end of the time interval $t'=(m+1)T$, the abundance of species $i$ is compared with the previous value. If $n_i(t')>n_i(t)$, the rewiring is accepted; otherwise the previous link $ij$ is restored \cite{Suweis2013}. The rewiring probability for the link $ij$ is $p_{ij}=1-k_j^{-\eta}$ ($\eta>0$), with $k_j$ being the degree of the partner species, so that a species with a lower degree is prone to keeping its link(s). This 
guarantees that any species interacts with at least one mutualistic partner species.

\section*{Acknowledgement}

This work was supported in part by U.S. Army Research Office under MURI Award No. W911NF-13-1-0340 and Cooperative Agreement No. W911NF-09-2-0053, by NSF grant DMS-1817124, and by DARPA award W911NF-17-1-0077.





\bibliographystyle{model1-num-names}
\bibliography{sample.bib}







\end{document}


\begin{frontmatter}


\title{Supplementary Information:\\A Dynamic Niche Model for the Emergence and Evolution of Mutualistic Network Structures}



\author{W. Cai, J. Snyder, A. Hastings and R. M. D'Souza}





\end{frontmatter}



We present here further analyses of the mutualistic networks in the framework of the dynamic niche model. The contents are organized as follows. We provide detailed definition and numerical implementation of the model in the first section. Static and dynamical properties of the evolved networks beyond those discussed in the main text are then addressed in the following two sections respectively. In the last part of this supplementary material, we discuss an extension of the original model that integrates the evolution of niche positions.

\section{Dynamic Niche Model: Definition and Numerical Simulation} \label{model_def}
We give here the definition of the niche overlap and updating rules of the dynamic niche model in extensive detail to complement the main text. We consider a mutualistic network consisted of $M_A$ animal species and $M_P$ plant species in two guilds, respectively. Each of the $M=M_A+M_P$ species is assigned a niche profile, which is characterized by a Gaussian function 
\begin{eqnarray}
     \label{eq:niche_overlap}
     H_i(s)=\frac 1 {\sqrt{2\pi}\sigma}e^{-(s-\overline{s_i})^2/2\sigma^2}
\end{eqnarray}
which is interpreted as the probability density of occupying the position $s$ on the niche axis \cite{MacArthur1967}. The niche centers $\overline{s_i}$, as the mean niche positions, are randomly dispersed on the niche axis $[0,1]$ according to a uniform distribution. We adopt a uniform niche width $\sigma$ for simplicity. We define the niche overlap $H_{ij}$ for a pair of species either within the same guild or across the guilds as
\begin{equation}
\label{eq:h}
H_{ij}=\int H_i(s)H_j(s)ds=e^{-(\overline{s_i}-\overline{s_j})^2/4\sigma^2}.
\end{equation}
The niche overlap can then be interpreted as the total joint probability of occupying the same position on the niche axis. 

Each species is involved in the cross-guild mutualistic interactions with selected partner species, while it competes with all rival species within its own guild. We represent the coupling relation by the matrix $\gamma^{AP}$ for mutualistic relations and $\beta^A$ or $\beta^P$ for competitive relations in respective guild $A$ or $P$. For either type of interaction, we assume that the coupling strength is proportional to the niche overlap of two species
\begin{subequations}
\label{eq:coupling}
\begin{align}
     \text{mutualistic:}\quad \gamma_{ik}&=\Omega_m\cdot \theta_{ik}\cdot H_{ik}\\
     \text{competitive:}\quad \beta_{ij}&=
     \begin{cases}
     1, &i = j\\
     \Omega_c\cdot H_{ij}, &i\not=j     
     \end{cases}     
\end{align}
\end{subequations}
where $i,j\in G=A$ or $P$ and $k\in \bar G=P$ or $A$. The coefficients of proportionality $\Omega_m$ and $\Omega_c$ are the interaction factors for mutualistic and competitive interactions, respectively, which represent the interaction strength per unit overlap. By changing these two factors, the overall interaction intensity can be controlled externally. $\{\theta_{ik}\}$ is the adjacency matrix:  $\theta_{ik}=1$ if species $i$ and $k$ interact, and $0$ if not. \\

\textbf{Updating rules} In the initial state, an uniform abundance is assigned to all $M$ species $n_i=n_0$ at $t=0$. For cross-guild mutualistic interactions, species are randomly connected across the guilds with a connectance $C_0$. More concretely, for each pair of species $i$ and $j$ ($i\in [1,M_A]$, $j\in [1,M_P]$), a random number $r$ is generated uniformly in $[0,1]$. A connection is formed if $r>C_0$ and left absent otherwise. For within-guild competitive interactions, we assume each species interacts with all species within the same guild. The coupling strengths are determined by the niche overlaps (Eq. \ref{eq:coupling}). At constant time intervals, species rewire repeatedly to change their niche relations in attempt to maximize their individual fitness. We adapt the rewiring process of \cite{Suweis2013} to the niche network. The following adaptation rules are executed. 

I. Rewiring: At the beginning of each time interval $t=mT$ ($m$ is a positive integer), we randomly select a species $i$ and one of its existing links $\gamma_{ij}$ is rewired to a randomly selected different mutualistic partner species $j'$ with probability $p_{ij}$: $\gamma_{ij}\longrightarrow\gamma_{ij'}$. When a new partner is connected, we evaluate the new mutualistic factor according to the niche overlap of the new species pair, that is, $\gamma_{ij'}=\Omega_m\cdot H_{ij'}$ (Eq. \ref{eq:coupling}a), and set $\gamma_{ij}=0$.

The rewiring probability is set to be $p_{ij}=1-k_j^{-\eta}$ ($\eta>0$). As such, species with a large number of partners are tolerant in losing links while species with a small number of partners are prone to keeping them. By using this condition, all participating species are guaranteed to have at least one partner, so that the connectance (total number of links) is constant over time with respect to the same number of total species.  

II. Population dynamics: After the rewiring, we allow the abundances of all species to settle to a new equilibrium, according to the generalized Lotka-Volterra equations with Holling-Type II mutualistic functional response \cite{Bastolla2009,Wright1989}
\begin{subequations}
\label{eq:lv}
\begin{eqnarray}
\frac{d n_i^A}{d t} &=& n_i^A\left(\rho_i^A  - \sum\limits_j\beta_{ij}^A n_j^A + \frac{\sum\limits_k\gamma_{ik}^{AP}n_k^P}{1 + h\sum\limits_k \theta_{ik}^{AP}n_k^P}\right)\\
\frac{d n_i^P}{d t} &=& n_i^P\left(\rho_i^P  - \sum\limits_j\beta_{ij}^P n_j^P + \frac{\sum\limits_k\gamma_{ik}^{PA}n_k^A}{1 + h\sum\limits_k \theta_{ik}^{PA}n_k^A}\right)
\end{eqnarray}
\end{subequations}
where $\rho_i$ is the intrinsic growth rate, $\{\gamma_{ij}\}$ and $\{\beta_{ij}\}$ are coupling strengths proportional to the niche overlaps (Eq. \ref{eq:coupling}), and $h$ is the handling time. The time interval $T$ is set sufficiently large to guarantee the dynamics to reach an equilibrium. Alternatively, one can integrate the dynamics until the variations of all abundance $\{n_i\}$ are limited within a sufficiently narrow window. 

III. Link recovery. At the end of the time interval $t'=(m+1)T$, we compare the current abundance of species $i$ to the previous value. If $n_i(t')>n_i(t)$, we keep the new link $\gamma_{ij'}$ as is; otherwise, we recover the link $\gamma_{ij'}\longrightarrow\gamma_{ij}$. The competitive factors $\beta_{ij'}$ do not change over time, since we consider here fixed niche positions (also see Sec. \ref{niche_evolution} where both interaction factors evolve).\\

\begin{figure*}[t]
\hskip -0.6cm
\includegraphics[width=1.055\textwidth]{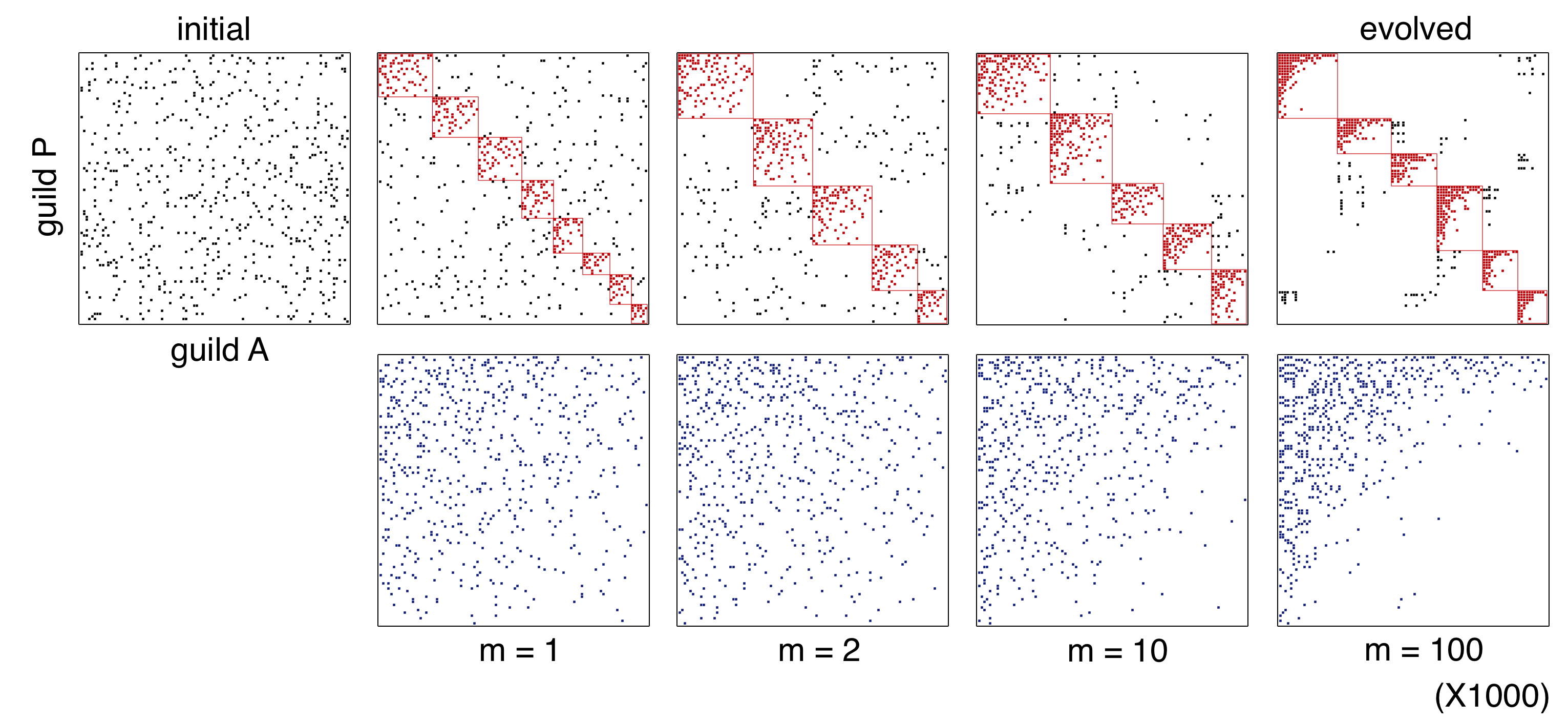}
  \caption{Emergence of network structures. Starting from an initial random bipartite network of $M_A=100$ animal species and $M_P=100$ plant species, both modular (upper panel) and nested (lower panel) structures are established through the evolution of niche relations. The snapshots of temporal link patterns are illustrated by the adjacency matrices at different times ($m=1000, 2000, 10000$ and $10^5$). The matrices are sorted by the modules and then by degrees within them for showing the modular structure, and simply by degrees for showing the nested structure, respectively. We mark the entries within the partitioned modules in red and those lying outside in black. The links are gradually absorbed into the modules (blocks in the upper panels) as the species are attaching to the local generalist hubs (upper left corners of the blocks). The evolved network at the steady state is significantly more modular and nested ($Q=0.6207$ and $N=0.9486$ for this example) than the randomized networks ($P<0.0001$). The connectance is fixed at $C_0=0.058$. A uniform growth rate $\rho=1$ is set for all species and the handling time $h=0.1$. The initial abundances are uniformly set to be $n_0=0.2$. }\label{fig_snap}
\end{figure*}

\textbf{Numerical simulations} The initial random network evolves into both highly modular and nested structures by following the above dynamical process, as shown in Fig. \ref{fig_snap}. We simulate the system for a sufficiently large time span, so that all macroscopic structural and populational parameters are able to settle into constant values within the simulation time, as shown in Fig. \ref{fig_tc}. This state is termed an ``evolved steady state". It means that the variations of the measures are consistently within a narrow bound over time. In this state, most of the rewiring attempts are rejected. 

The level of nestedness is measured in terms of the nestedness temperature coefficient (NTC), which approaches $1$ when the network is perfectly nested and decreases when the network is less nested \cite{Rodriguez-Girones2006}. The level of modularity of the adjacency matrix ${\theta_{ij}}$ is calculated using Newman's leading eigenvector algorithm \cite{Newman2006}, which seeks a partition of the network that maximizes the modularity quality function 
\begin{equation}
\label{eq:modularity}
Q=\frac 1 {2L} \sum_{i,j} \left(\theta_{ij}-\frac{k_ik_j}{2L}\right)\delta(c_i,c_j)
\end{equation}
where $k_i$ is the node degree, $2L$ is the total number of links, and $c_i$ is the module that node $i$ belongs to under a certain partition. We used the tool package BiMat for the calculations of the two structural measures \cite{Flores2016}. 

In all simulations, we set the time interval $T=100$. A uniform growth rate $\rho=1$ is used for simplicity. The connectance is chosen to be $C_0=0.058$, which is close to the average value of the empirical networks containing $M=200$ species \cite{Suweis2013}. We set the exponent $\eta=1$; we find that simulation results are insensitive to the choice of $\eta$ in a large range ($\eta\in[1,+\infty)$). The handling time $h$ determines the saturation level in the functional response \cite{Wright1989}, which is set to be $0.1$ for Fig. 1, 2 and 4, and $0.5$ for Fig. 3 in the main text. The properties remain qualitatively unchanged for the choice of $h$. The value 0.5 for Fig. 3 was chosen to highlight the crossover regime. The scatter plots in Fig. 2c and 2d in the main text consists of 400 realizations, which are generated for randomly chosen interaction factors $\Omega_m$ and $\Omega_c$ in the range $\left[0.01,0.15\right] \times \left[0.01,0.15\right]$. 

To avoid edge effects, we define the niche axis to be periodic so that species are roughly symmetric on the niche axis: each species has approximately equal numbers of rival species in the same guild and potential mutualistic partner species in the opposite guild. If a fixed niche boundary condition is used, the simulation results are qualitatively similar. The niche centers are still confined within the interval $\left[0,1\right]$, but the niche overlaps are calculated on the one-dimensional axis without periodic condition \cite{Scheffer2006}. In such cases, species with niche centers close to the opposite boundaries are unlikely to be partitioned in the same module. \\ \\

\section{Static Properties of Evolved Networks} \label{static}
Numeric simulations show that all macroscopic structural and demographic measures settle to constants asymptotically (Fig. \ref{fig_tc}b and \ref{fig_tc}c). Most rewiring attempts are rejected at the evolved steady state, which keep fluctuations of these measures within a narrow range. The clustering process undergoes a transitory period and evolves into a relatively fixed number of modules. Only a small proportion of specialists may still switch between neighbouring modules, while the interconnections among generalist partners across the guilds are preserved. In this section, we address more relations observed in the structural and demographic measures of the evolved mutualistic network at the steady state.

\subsection{Correlations at Evolved Steady State}
We show in the main text that the macroscopic measures at the evolved steady state are determined by local interaction factors. We thus check the relations of these measures of an ensemble of generated networks by randomly choosing interaction factors $\Omega_m$ and $\Omega_c$ while keeping the network size and connectance fixed. It is straightforward to see that the structural measures, modularity $Q$ and nestedness $N$, are negatively correlated, as shown in Fig. \ref{fig_mod_ntc_nd}a. This is simply due to the topological constraint: a higher number of modules of smaller sizes (higher modularity $Q$) tend to avoid overlaps of partnerships, which is expressed by a low nestedness $N$ by its definition.


The niche affinity is reflected in the evolved modular and nested structure, as long suggested by empirical studies \cite{Olesen2007,Santamaria2007,Guimaraes2007}. This is again demonstrated by generating networks for randomly chosen interaction factors $\Omega_m$ and $\Omega_c$. The average niche distance $\bar{d}$, over all connected species pairs, is found to be negatively correlated with the modularity $Q$ (linear fitting) and positively correlated with the nestedness $N$ (exponential fitting), as shown in Fig. \ref{fig_mod_ntc_nd}b. Hence, a more nested network structure can tolerate interspecific partnerships with less niche affinity (higher $\bar{d}$), while a more modular structure (smaller sizes of modules) is packed with species of more complementary niches (lower $\bar{d}$).

\subsection{Bounded structure-population relation} The adaptation process is facilitated by a positive feedback between the structural and demographic heterogeneities. Consequently, the structural measures of an evolved network are manifestations of the demographic heterogeneity in species abundances. However, no coherent relation exists between either modularity or nestedness and the overall population of the community $\Sigma_i n_i$ (equivalently the mean abundance per species $\bar{n}$). Rather, the structural measures are approximately bounded against the mean abundance $\bar{n}$ (Fig. \ref{fig_ntc_mod_abd}), suggesting that relatively high nestedness or low modularity is achievable only when the overall population is low. The relation is demonstrated for an ensemble of generated networks for randomly chosen interaction factors. This implies that although the mutualistic network structure is formed by maximizing individual abundances, being more nested or modular does not contribute monotonically to the overall population. The previous claim that greater nestedness facilitates greater population is only valid under special conditions \cite{Suweis2013}.

\subsection{Role of Niche Width} 
We have analyzed the structural properties by specifying a niche width $\sigma$, which can be interpreted as the occupation probability on the niche axis. For simplicity, a uniform value is used to represent the average niche width. Similar as the interaction factor, this parameter also regulates the interspecific interaction pattern since it is directly associated with the average number of cross-guild partners. The sizes of modules increase with the niche width and thus allow a more hierarchical structure within individual modules. Consequently, the nestedness and modularity exhibit a positive and negative correlation with the niche width $\sigma$, respectively, as shown in Fig. \ref{fig_nq_nw}a. We will further analyze its impact on the dynamical properties in the following section.\\ \\

\section{Dynamical Properties of Evolved Networks}  \label{dynamical}
We now examine dynamical properties of mutualistic networks in the framework of structural adaptation. We are most concerned with the network resilience, which reflects to what extent the obtained static properties are maintained under external influence. As discussed in the main text, we consider influences at both small and large time scales.

\subsection{Ecological Time Scale} 
We assume that the interaction factors do not vary during a transitory perturbation, so that the network topology remains fixed at the evolved steady state. Such perturbation is associated with the ever-existing small stochastic environmental changes in the species abundances. This allows us to use the real part of the leading eigenvalue of the Jacobian matrix of the network dynamics Eq. \ref{eq:lv} as the measure of the local stability, that is, $S=-Re(\lambda)_{max}$. This value is always calculated at the end of the time interval $T$ when the network settles to the equilibrium. Since most rewiring attempts are rejected for the evolved network, the stability measure $S$ only fluctuates around a constant value with a small variation (see Fig. \ref{fig_tc}d).

\textbf{Local stability} At equilibrium, the Jacobian $\Phi$ adopts a relatively simpler form, since the right handside of Eq. \ref{eq:lv} equals zero.
\begin{eqnarray}
     \label{eq:jacob_def}
     \Phi=
  \begin{bmatrix}
    \Phi^{AA} & \Phi^{AP} \\
    \Phi^{PA} & \Phi^{PP} 
  \end{bmatrix}
  =
  \begin{bmatrix}
    \partial\dot{n_A}/\partial n_A & \partial\dot{n_A}/\partial n_P \\
    \partial\dot{n_P}/\partial n_A & \partial\dot{n_P}/\partial n_P 
  \end{bmatrix}
\end{eqnarray}
where the components can be written as (symmetric for $\Phi^{PP}$ and $\Phi^{PA}$)
\begin{eqnarray}
     \label{eq:jacob_aa_ap}
     \Phi^{AA}&=&-\text{diag}(n_A)\cdot\beta^A \\
     \Phi^{AP}&=&-\text{diag}(n_A)\cdot\xi^{AP}
\end{eqnarray}
with
\begin{eqnarray}
     \label{eq:xi_ap}
     \xi^{AP}_{ij}=\frac{\left[\gamma^{AP}+h\cdot \text{diag}(\rho^A-\beta^A n_A)\theta^{AP}\right]_{ij}}{\left(1+h\cdot \left[\theta^{AP}n_P\right]_i\right)}
\end{eqnarray}
The symbol $\text{diag}(v)$ denotes an $L\times L$ diagonal matrix whose diagonal entries are the elements of the vector $v\in \mathbb{R}^L$. We have nummerically calculated the eigenvalues for all figures, according to Eq. \ref{eq:jacob_def} - \ref{eq:xi_ap}. However, following the perturbation expansion in \cite{Suweis2013}, in the limit when both $\gamma_{ij}$ and $\beta_{ij}$ $(i\not= j)$ are far smaller than the diagonal entries $\beta_{ii}$ (self-competition coefficient), the eigenvalue of the Jacobian is predominantly determined by the abundances at equalibrium, that is, $\lambda_i\approx -n_i\cdot\beta_{ii}$. Hence, the stability measure is approximately only related to the minimum abundance of the poorest species in the community $S\approx -n_{min}\cdot\beta_{ii}$ (we set $\beta_{ii}=1$). This simple relation links the stability of the community and the lower bound of the population (see inset of Fig. \ref{fig_stab_mod}a).

\textbf{Bounded relation between stability and structures.} The network structure has a systematic impact on the network stability. The question how nestedness or modularity could influence the stability has aroused substantial interest. In the framework of our dynamic model, however, the stability $S$ shows only a bounded positive and negative tendency with modularity $Q$ and nestedness $N$, respectively (see Fig. \ref{fig_stab_mod}). This is demonstrated for generated networks corresponding to randomly chosen interaction factors $\Omega_m$ and $\Omega_c$ (Fig. \ref{fig_stab_mod}). Modularity $Q$ or nestedness $N$ is correlated with the demographic heterogeneity $CV(n)$ of the species abundances, but still allows a broad range of the lower bound of abundance $n_{min}$. A simple correlation of the stability and any structural measure is thus absent.

\textbf{Critical role of competition intensity.} In the main text, we identify the crucial role of competitive interaction on stability. Its intensity determines whether enhancing mutualistic interaction by the factor $\Omega_m$ would stabilize or destabilize the network. Contrasting tendencies are correspondingly revealed in the demographic distribution (Fig. \ref{fig_abd_dist}): For $\Omega_c$ below the transition point $\Omega_c^S$, the entire distribution shifts towards greater abundance with increasing $\Omega_m$; for $\Omega_c$ above $\Omega_c^S$, the average abundance increases slightly but the distribution becomes much broader, with both highest and lowest limits extending in the opposite directions. This is consistent with the tendencies of the stability through the relation between the minimum abundance $n_{min}$ and stability $S$. The demographic distribution becomes bi-modal at high $\Omega_m$ for $\Omega_c>\Omega_c^S$, which indicates that the increase of abundances in the rich species is associated with the decrease in the poor species. 

In contrast, the relation between stability and niche width is independent of the competitive interaction factor $\Omega_c$ (see Fig. \ref{fig_nq_nw}c and \ref{fig_nq_nw}d). This is again consistent with the lower bound $n_{min}$ of the species abundances. In fact, the entire demographic distribution shifts persistently lower with the niche width $\sigma$ for any competitive interaction factor $\Omega_c$ (Fig. \ref{fig_nq_nw}b). This tendency suggests that the mutualistic network is always destabilized when species may interact with partners within a broader range of niches on average.

\subsection{Evolutionary Time Scale}
Local interactions, in terms of their intensities and specificity of relations, can alter with the environment at the evolutionary time scale. We show in the main text the adaptation of the interspecific relation and demographic distribution. To examine if the original state of a mutualistic network is retrievable under large environmental changes, we trace an adaptive process by gradually raising and then reversing the mutualistic interaction factor $\Omega_m$. We use a sufficiently slow change rate, $\Delta\Omega_m=10^{-6}$ per time interval, so that the network is always at a quasi-evolved-steady-state for all $\Omega_m$ values.

The trajectory of either modularity $Q$ or nestedness $N$ shows non-overlapping ascending and descending paths of $\Omega_m$. Similar hysteretic adaptation of network structure is also observed by continuously changing the niche width $\sigma$, as shown in Fig. \ref{fig_irv_nw}. This irreversibility originates from the broken symmetry in the merging and splitting processes of modules by broadening and narrowing the niche width $\sigma$. Species show resistance when detaching from the highly abundant generalists, which comprise the cores of the modules. Such asymmetry however does not affect either the average abundance $\bar{n}$ or the network's local stability $S$ (see insets of Fig. \ref{fig_irv_nw}). Hence, cooperative species may exhibit alternative relations for the same local interaction elements (interaction factor or niche width), while such alternative stable states do not change the local stability or the overall community population.

\subsection{Cumulative Advantage and Extinction}
All species are subject to competition. Thus, the entire demographic distribution shifts downward due to enhanced competition, which is responsible for potential extinctions \ref{fig_abd_dist}a \cite{Dakos2014,Jiang2018}. In contrast, enhancing the intensity of mutualistic interaction tends to raise the overall community population. However, the demographic distribution shows that the abundances of the rich and poor species change in opposite directions (Fig. \ref{fig_abd_dist}b). Excessive mutualistic interaction may thus cause extinction, due to such cumulative advantage: the rich species become even richer along with the loss of the poor species. 

By continuously increasing the mutualistic interaction factor $\Omega_m$, we find that the least-abundant species becomes extinct beyond a threshold $\Omega_m^{ex}$. The number of species drops drastically in a narrow range of $\Omega_m$, while the network becomes less nested (its modularity does not vary strongly), as demonstrated in Fig. \ref{fig_ex}a. Along with the structural change, the average abundance per remaining species increases more rapidly than before the extinction (Fig. \ref{fig_ex}b), simply due to the deletion of the low-abundant species. The simulated extinction reveals the potential negative effect of an increasingly hierarchical community with enhanced mutualism. It roots in the fact that the poor-poor coalition is not encouraged in the adaptive process, which is targeted at maximizing the utilization of resources. \\

\section{Niche Evolution and Limiting Similarity}  \label{niche_evolution}

The adaptive dynamics of mutualistic network may involve the evolutions of niche relation and niche positions simultaneously at a long evolutionary time scale. They can be incorporated in an extension of our model: a randomly chosen species is allowed to change its niche position (the niche center $\bar{s_i}$) by a small shift $\Delta s$ on the niche axis with a small probability $p_t$ at constant time intervals, in addition to the rewiring with probability $1-p_t$. The shift can be in the positive or negative direction according to a binary random number. If the change in the niche position contributes positively to the individual abundance at the end of the time interval, the new niche position $\bar{s_i}'$ is accepted; otherwise the niche is reversed to the previous position $\bar{s_i}$. 

The numerical simulations show that starting from an ensemble of niches that are randomly scattered on the niche axis according to a uniform distribution, all niches converge to a stable distribution on the niche axis that contains multiple narrowly confined lumps at the evolved steady state, as shown in Fig. \ref{fig_tr_time} and Fig. \ref{fig_tr}a. Such limiting similarity of niches has long been observed in the classical model of pure niche competition (in the framework of invasive fitness) \cite{MacArthur1967,Scheffer2006}. However, the lumps of niches in the evolved mutualistic network are even more confined, showing good separation from the neighbouring lumps. The network is partitioned into a number of modules that correspond to these lumps, each containing a nested structure (Fig. \ref{fig_tr}b and \ref{fig_tr}d). Due to the convergence of niche positions, the network shows consistently higher modularity than that in the original model defined in Sec. 1. The demographic and degree distributions are hierarchical within each lump (Fig. \ref{fig_tr}c).

The niche convergence is exhibited in a large range of nonzero $p_t>0$, even for very small probabilities. The convergence is guided by the effective suppression of common competitors, as elucidated in \cite{MacArthur1967,Scheffer2006}. All competition interactions are intensified by the mutualistic interactions, which tend to increase both abundances of the rivalry species. Alternatively, we can start the evolution from the initial condition that all species are located at the same niche position $\bar{s_i}=0.5$. In this case, the abundances will first decrease to zero in the transitory period, simple due to that all species are at the same position and compete strongly. We thus need to add a very small constant term ($\mu=10^{-6}$) at the right hand side of Eq. \ref{eq:lv}, to guarantee that any species may recover from a tiny abundance. The numerical simulation shows again that the species spread on the niche axis and converge to a steady lumped niche distribution with a nested and modular structure (similar to that shown in Fig. \ref{fig_tr}). The latter process demonstrates clearly how the heterogeneous network structure can be established from a completely homogeneous state as a consequence of niche adaptation.

\bibliographystyle{model1-num-names}
\bibliography{sample.bib}

\newpage

\begin{figure*}[t]
\hskip -0.7cm
\includegraphics[width=1.06\textwidth]{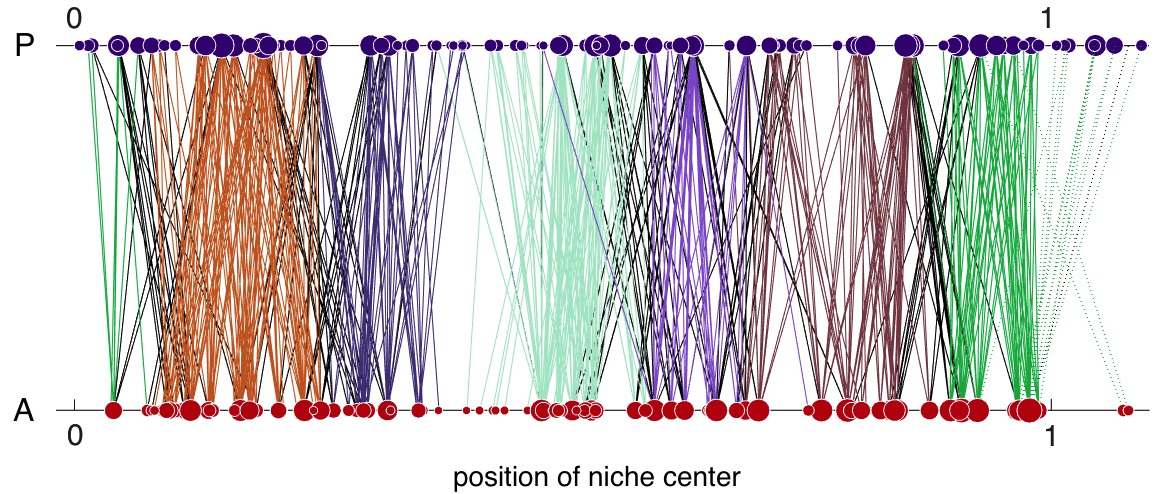}
  \caption{Evolved link pattern and distribution of abundances. Links belonging to different modules are marked in colours, according to the partition of the final snapshot in Fig. \ref{fig_snap}. The modules are connected by the black-coloured links lying outside all modules. The dotted lines are links that wind around the periodic boundary (nodes close to the right side of $s=0$ are duplicated at $s>1$). The species abundances, as represented by the sizes of the disks, show a heterogeneous demographic distribution along the niche axis. }\label{fig_pattern}
\end{figure*}

\begin{figure*}
\hskip -0.75cm
  \includegraphics[width=1.07\textwidth]{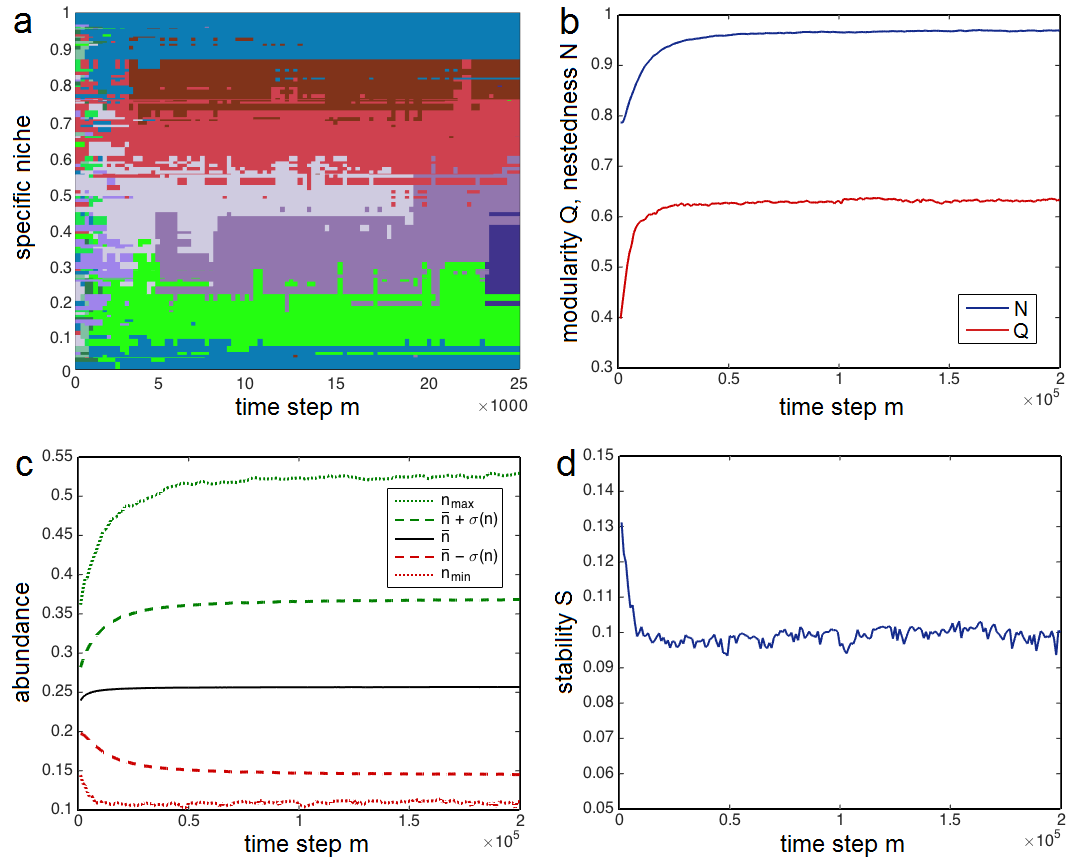}
  \caption{Evolutionary time courses to steady states. \textbf{a}, Evolving network partition. Each colour represents a module, in which species are more densely connected internally than to those in other modules. The number of modules decreases rapidly over the transitory period and stabilizes at the evolved steady state. \textbf{b}, Structural measures. Both modularity $Q$ and nestedness $N$ reach a steady state after a transitory time period. Each time interval lasts for a fixed integration time $T=100$. \textbf{c}, Time courses of abundance measures, showing the mean abundance per species $\bar{n}$, the deviation $\sigma (n)$ over the community, and the lower and upper bounds $n_{min}$ and $n_{max}$. \textbf{d}, Time course of local stability. The network stability $S=-Re(\lambda)_{max}$ is measured by the real part of the leading eigenvalue of the Jacobian matrix of Eq. \ref{eq:lv} at the end of every time interval, when the population have settled to an equilibrium. $M_A=100$ animal species and $M_P=100$ plant species are involved in the simulations. The curves in panels \textbf{b} to \textbf{d} are averaged over 10 realizations.}\label{fig_tc}
\end{figure*}
\begin{figure*}[t]
\hskip -0.65cm
  \includegraphics[width=1.08\textwidth]{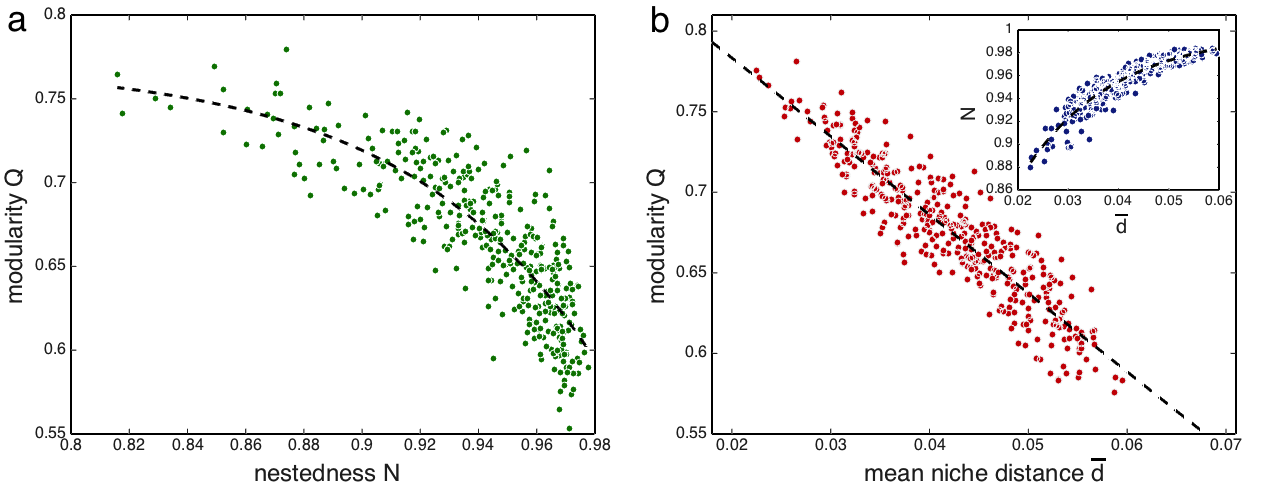}
  \caption{Evolved network structure. \textbf{a}, Correlation of structural measures. Modularity $Q$ and nestedness $N$ are negatively correlated due to the topological constraint: modules of smaller sizes tend to prevent overlaps of partners of species in different modules. The scattered data points are fitted by an exponential curve. \textbf{b}, Modularity and nestedness (in inset) versus average niche distance $\bar{d}$ of linked species. Modularity decreases linearly while nestedness increases exponentially with $\bar{d}$. The networks are obtained for 400 pairs of randomly chosen interaction factors $\Omega_m$ and $\Omega_c$ according to a uniform distribution in $[0.01,0.15]\times[0.01,0.15]$. }\label{fig_mod_ntc_nd}
\end{figure*}

\clearpage
\begin{figure*}
\hskip -0.9cm
  \includegraphics[width=1.07\textwidth]{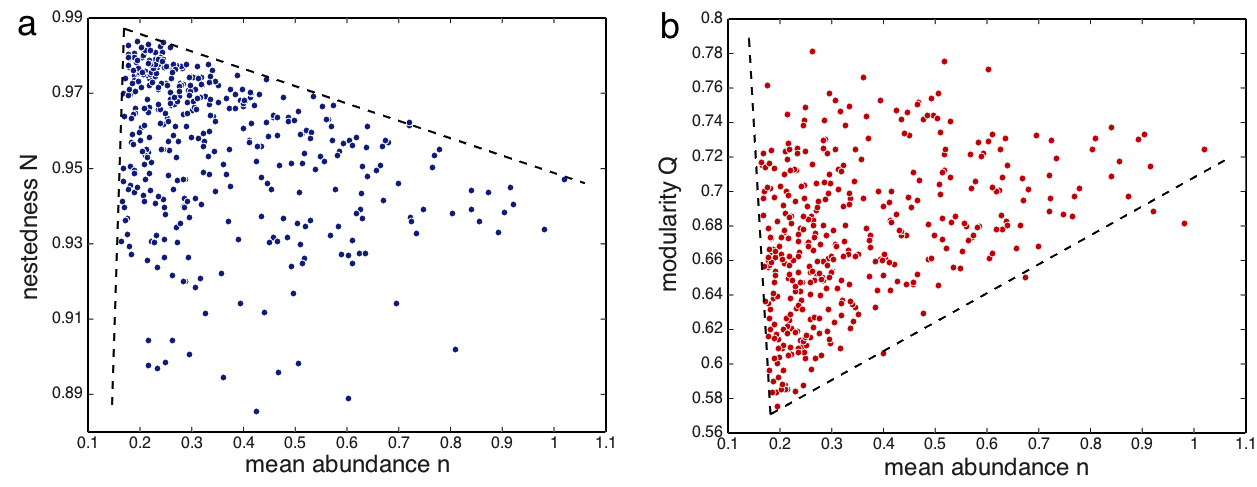}
  \caption{Bounded relation of network structure versus community population. Structures with relatively high nestedness (panel a) and low modularity (panel b) can be reached only in the region of low mean abundances $\bar{n}$. All data points are obtained by varying the interaction factors randomly according to a uniform distribution as for Fig. \ref{fig_mod_ntc_nd}.}\label{fig_ntc_mod_abd}
\end{figure*}
\begin{figure*}[h]
\hskip -0.8cm
  \includegraphics[width=1.06\textwidth]{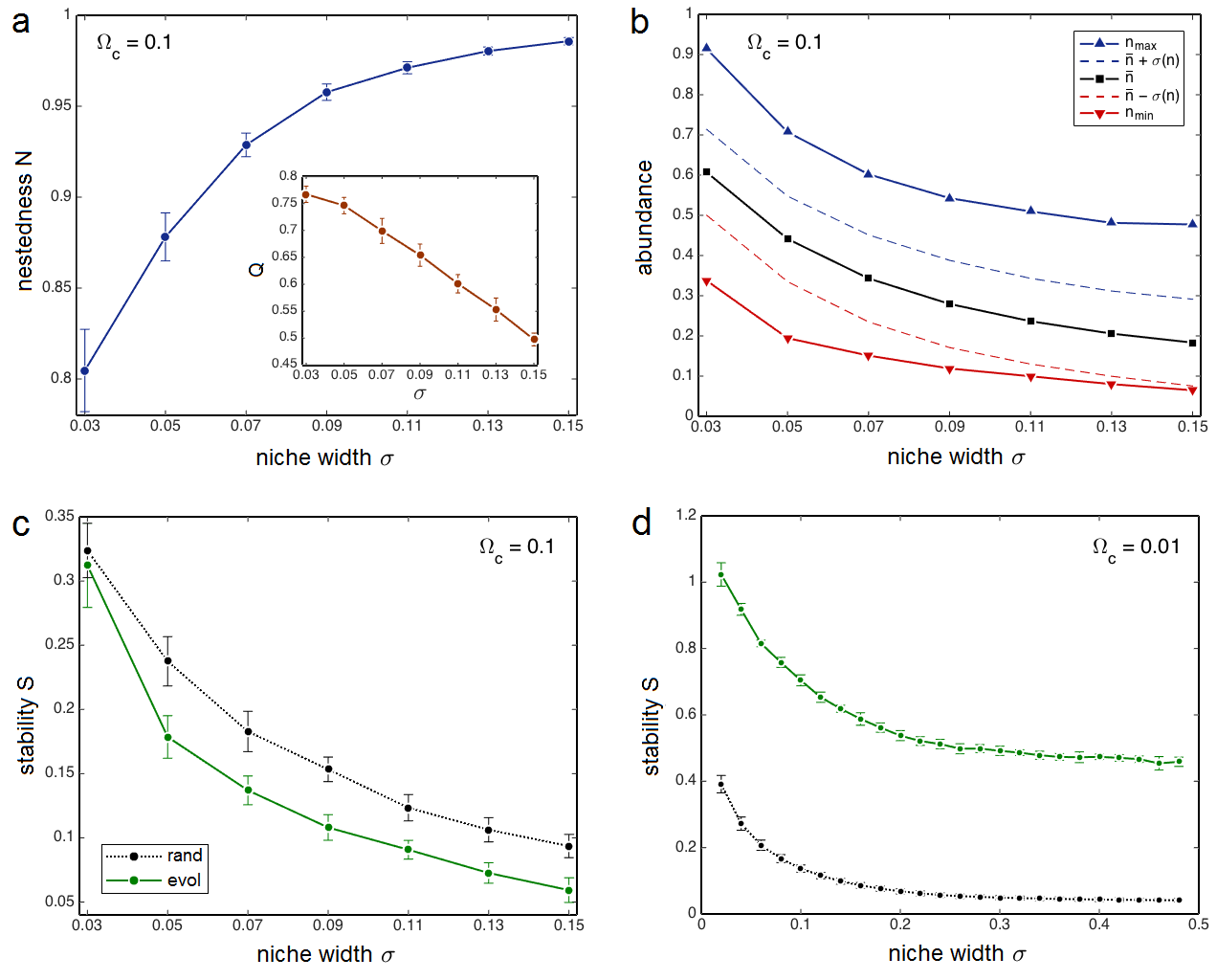}
  \caption{Role of niche width. \textbf{a}, Structural change. The nestedness and modularity (inset) exhibit a positive and negative correlation with the niche width $\sigma$, respectively. For simplicity, all species are assigned the same niche width. \textbf{b}, Monotonic change in population. For any competition factor $\Omega_c$, broadening the niche width always shifts down the entire demographic distribution, including the overall population and the lower and higher bounds of the community. We show here the case for $\Omega_c=0.1$, but similar tendencies are exhibited for other values of $\Omega_c$. \textbf{c, d}, The local stability $S$ consistently decreases with the niche width $\sigma$, irrespective of the competition factor $\Omega_c$. However, the competition factor still determines the relative stability: the network is consistently less and more stable than the randomized networks (dashed lines) for $\Omega_c=0.1$ and $0.01$, respectively. }\label{fig_nq_nw}
\end{figure*}
\begin{figure*}[h]
\hskip -0.8cm
  \includegraphics[width=1.07\textwidth]{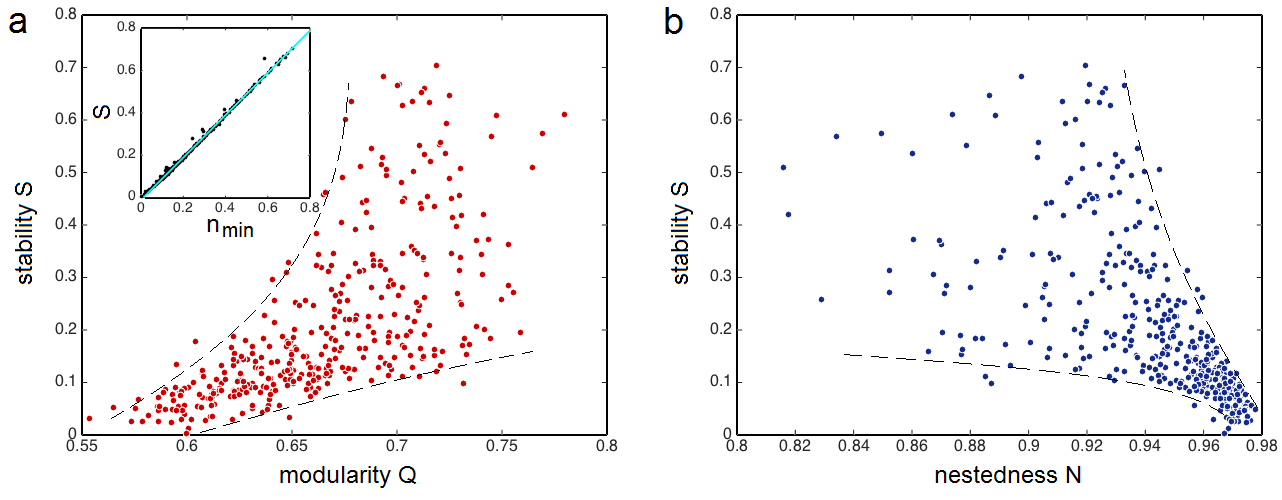}
  \caption{Bounded relation between stability and structure. The network stability $S$ shows only an approximately bounded positive and negative tendency with modularity $Q$ (panel \textbf{a}) and nestedness $N$ (panel \textbf{b}), respectively. Inset: The stability $S$ is correlated with the lower bound of the specific abundance over the community, through the simple relation $S\approx n_{min}$ (when all $\beta_{ii}=1$) when the interspecific interaction is far less intensive compared with the intraspecific competition. The networks analyzed here are randomly generated as for Fig. \ref{fig_mod_ntc_nd}.}\label{fig_stab_mod}
\end{figure*}
\clearpage
\begin{figure*}
\hskip -0.8cm
  \includegraphics[width=1.07\textwidth]{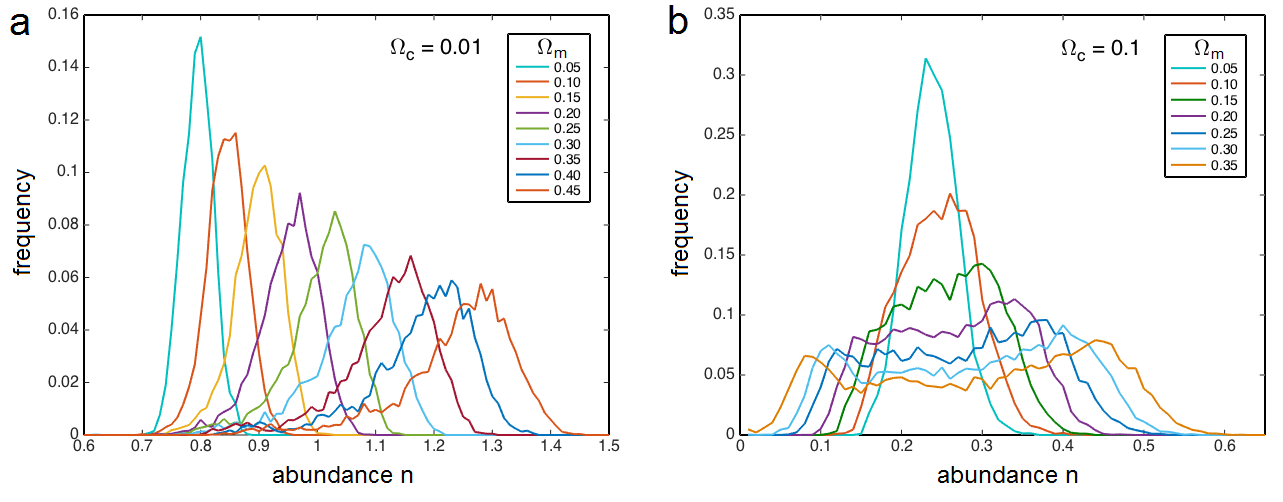}
  \caption{Contrasting modes of changes in demographic distribution. \textbf{a}, For the competitive factor $\Omega_c$ below the transition point $\Omega_c^S (\approx 0.02)$, the entire distribution shifts up with the mutualistic factor $\Omega_m$. \textbf{b}, For $\Omega_c > \Omega_c^S$, the distribution becomes substantially broader, with both highest and lowest limits extending in the opposite directions. In the latter case, the overall population of the community increases slightly. The demographic distribution tends to be bi-modal at high $\Omega_m$ for $\Omega_c>\Omega_c^S$, indicating that the increase of abundances in the rich species is associated with the decrease in the poor species. The handling time is set to be $h=0.5$. }\label{fig_abd_dist}
\end{figure*}
\begin{figure*}
\hskip -0.75cm
  \includegraphics[width=1.07\textwidth]{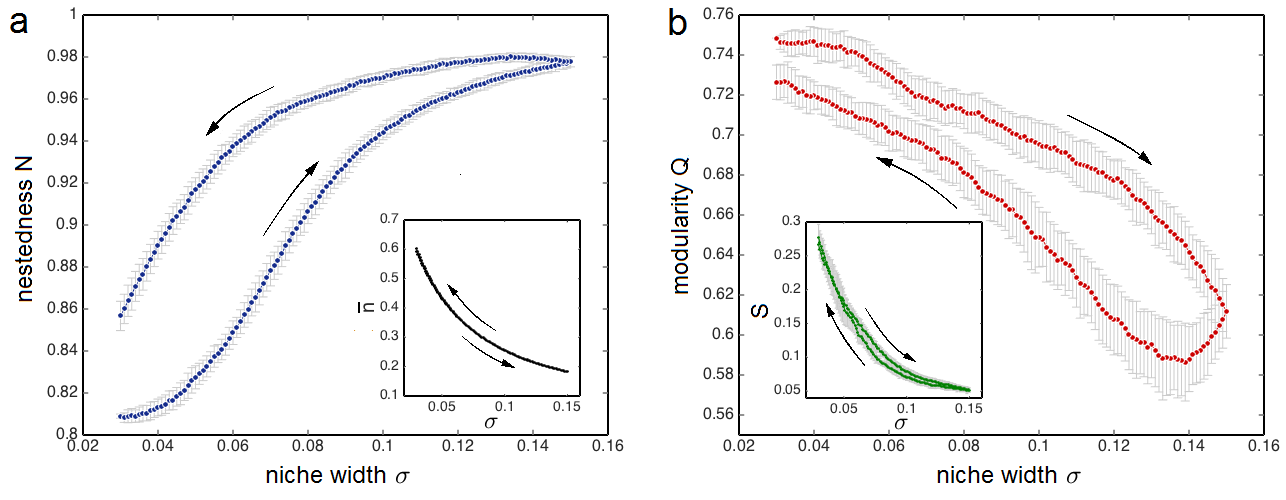}
  \caption{Hysteresis in the adaptation with niche width. Similar to the case for the mutualistic factor, the network structures (nestedness $N$ in panel \textbf{a} and modularity $Q$ in panel \textbf{b}) exhibit history-dependent paths with the niche width $\sigma$, which is raised and lowered by external control in the interval $[0.03, 0.15]$. A slow change rate, $\Delta \sigma =10^{-6}$ per time interval, is used to guarantee that the system remains at the quasi-steady-state during the adaptation. It shows again that the mutualistic network may be at alternative stable states under the same local interaction parameters. Insets: The mean abundance per species $\bar{n}$ and the stability measure $S$ however show no hysteresis with the control parameter $\sigma$ (the slight openness in the path of $S$, caused by fluctuations, is negligible). The trajectories are averaged over 50 realizations, where the error bars represent the standard deviations.}\label{fig_irv_nw}
\end{figure*}
\begin{figure*}
\hskip -0.6cm
  \includegraphics[width=1.09\textwidth]{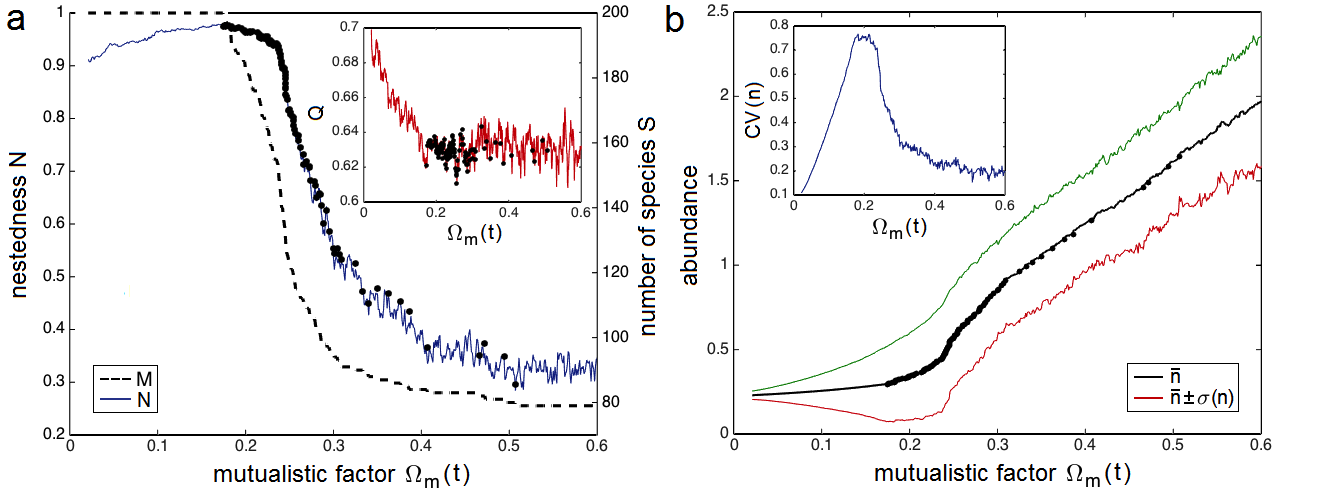}
  \caption{Cumulative advantage and extinction. \textbf{a}, By increasing the mutualistic factor $\Omega_m$ continuously, the least-abundant species becomes extinct beyond a critical point $\Omega_m^{ex}$ $(\approx 0.18)$. The number of species drops drastically in a narrow range of $\Omega_m$ (between 0.18 and 0.3). The nestedness $N$ of the network comprised of the remaining $M$ species decreases rapidly at the meantime, whereas its modularity $Q$ does not vary strongly (inset). Each black spot represents an extinction event of one species. \textbf{b}, The average abundance per remaining species $\bar{n}$ increases more rapidly than before the extinction. Inset: The demographic heterogeneity $CV(n)$ reduces with $\Omega_m$ beyond the critical point $\Omega_m^{ex}$. The change rate of $\Omega_m$ per time interval is set to be sufficiently small ($10^{-6}$) as in Fig. 4 of the main text. }\label{fig_ex}
\end{figure*}
\begin{figure*}
\hskip -0.78cm
  \includegraphics[width=1.07\textwidth]{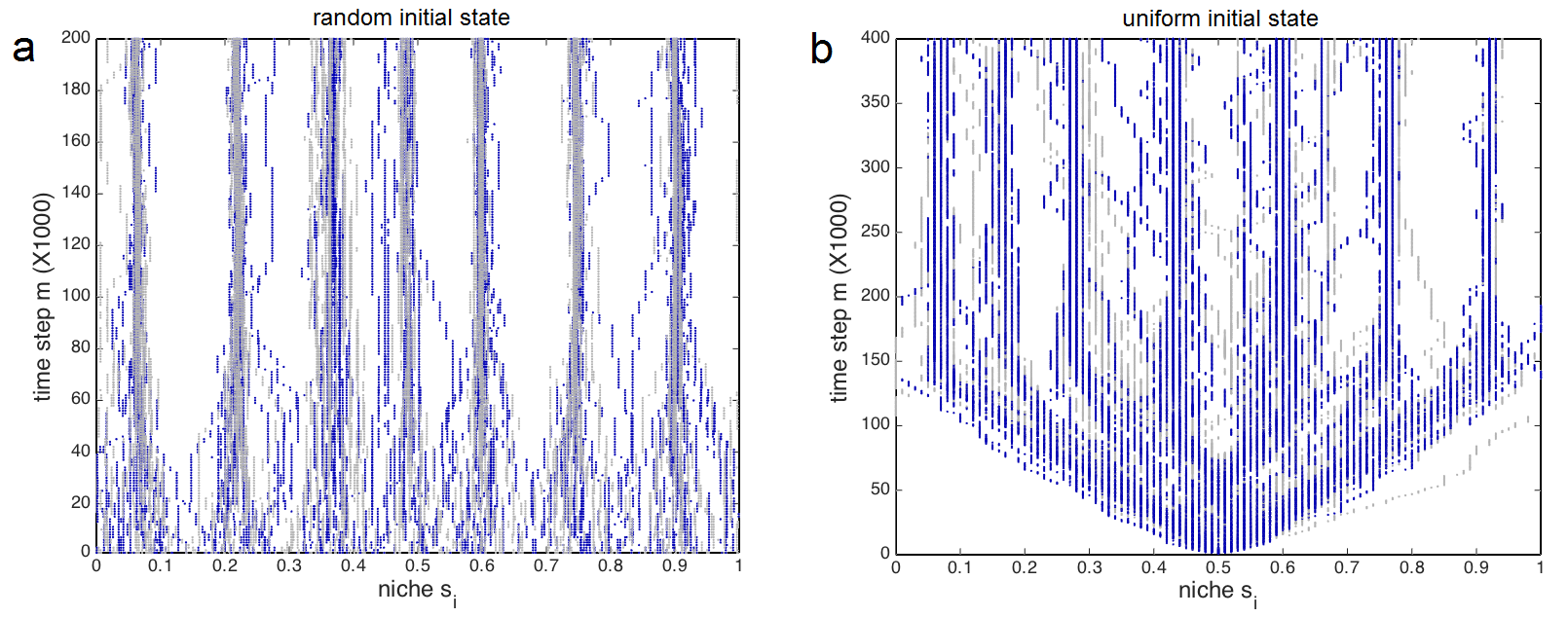}
  \caption{Evolution trajectories of niche positions. In the extension of the original model, the niche positions are allowed to shift with probability $p_t=0.1$ at fixed time intervals, in parallel to the cross-guild rewiring with probability $1-p_t=0.9$. The species evolve into a steady state with all niches converging to well constrained lumps. The niche positions of the two guilds are marked in blue and gray, respectively, which become aligned approximately in each lump. The evolution is simulated with two different initial conditions. \textbf{a}, Random initial condition: species are assigned with a niche position on the niche axis, with its center $\bar{s_i}$ chosen uniformly in $[0,1]$ by random. \textbf{b}, All species start from the same niche position ($\bar{s_i}=0.5$). The initial abundances are uniformly set to be $n_i(t=0)=0.2$. We set the distance of shift to be $\Delta s=0.01$ per time interval, but the final result is insensitive to the choice of $\Delta s$. }\label{fig_tr_time}
\end{figure*}
\begin{figure*}
\hskip -0.68cm
  \includegraphics[width=1.06\textwidth]{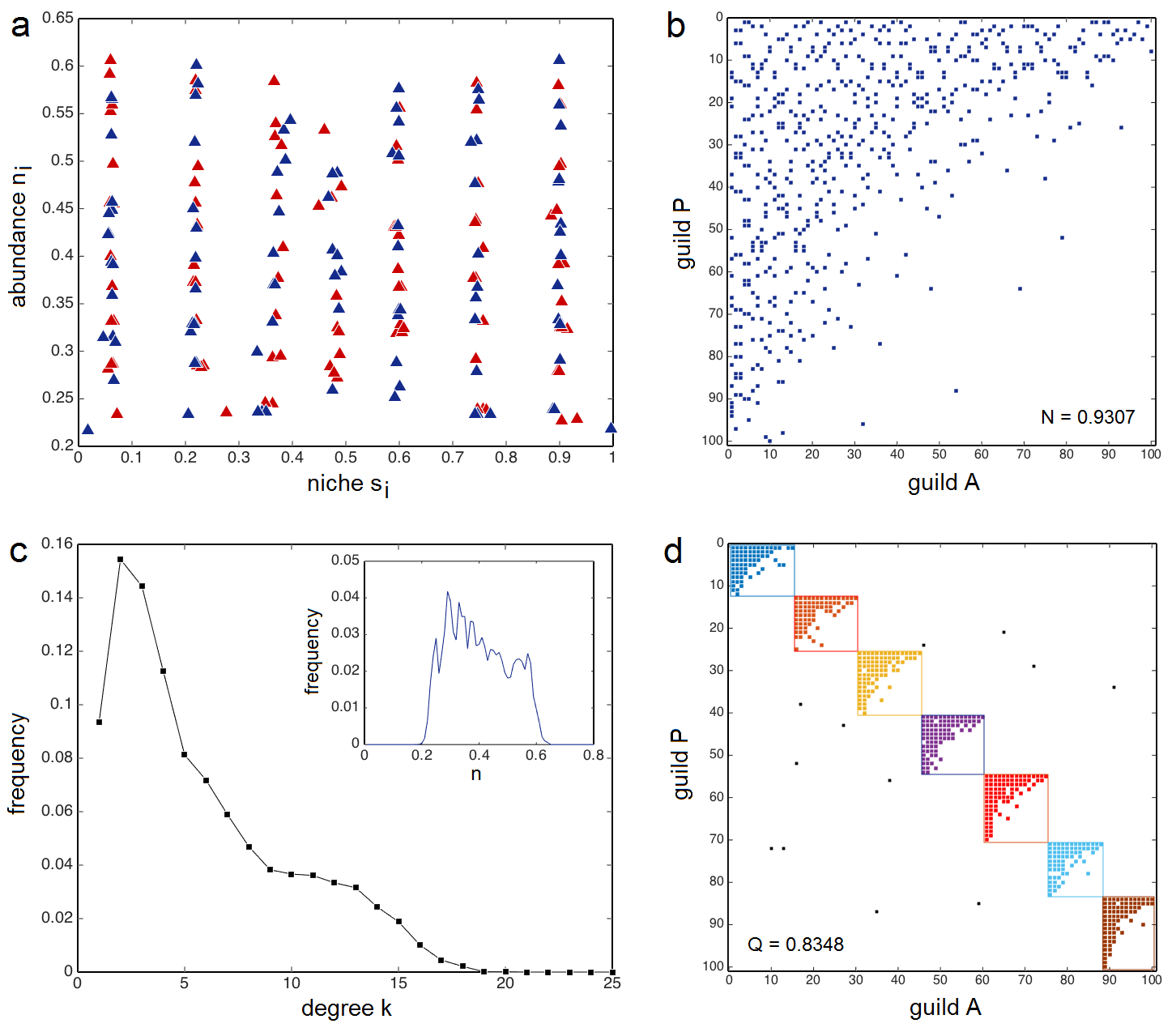}
  \caption{Niche convergence and limiting similarity. \textbf{a}, Lumped distribution of species abundances on the niche axis, for the extended model involving the adaptation of both niche linkage and positions. The abundances are represented by blue and red symbols for guild $A$ and $P$, respectively. \textbf{b}, The adjacency matrix showing the nested network structure at the evolved steady state, where the species are sorted by the degrees. \textbf{c}, The hierarchical degree and abundance (inset) distributions of the evolved network. \textbf{d}, The adjacency matrix for the same network, but sorted first by modules and then by the species degrees within each of them. Due to the niche convergence, the evolved network shows a relatively higher modularity $Q$ than the one in the original model. The random initial condition is applied, as for Fig. \ref{fig_tr_time}a. }\label{fig_tr}
\end{figure*}










